\documentclass{appolb}
\usepackage{graphicx}

\begin{document}
\title{The Brief Life of a Hadron: QCD unquenched%
\thanks{Presented at EEF70: Workshop on Unquenched Hadron Spectroscopy: Non-Perturbative Models and Methods of QCD vs. Experiment, (1-5 September, 2014) University of Coimbra, Portugal.}%
}
\author{Michael R. Pennington
\address{Theory Center, Thomas Jefferson National Accelerator Facility,\\ Suite 1, 12000 Jefferson Avenue, Newport News, VA 23606, U.S.A.}
\\
}
\maketitle
\begin{abstract}
Once upon a time, the picture of hadrons was of mesons made of a quark and an antiquark, and baryons of three quarks. Though hadrons heavier than the ground states inevitably decay by the strong interaction, the successes of the quark model might suggest their decays are a mere perturbation. However, Eef van Beveren, whose career we celebrate here, recognised that decays are an integral part of the life of a hadron. The channels into which they decay are often essential to their very existence. These hold the secrets of strong coupling QCD and teach us the way quarks really build hadrons. 
\end{abstract}
  
\section{The life of a hadron}
The study of the spectrum of hadrons is vital to improving our understanding of how strong coupling QCD really works: binding quarks (and antiquarks) into colour neutral objects, determining their individual properties and their collective behaviour in nuclei of which we and the visible universe are made. The octet of lightest baryons: the proton and neutron, and their strange and stranger cousins, are all stable as far as the strong interaction is concerned. On time scales of the strong interaction of 10$^{-23}$ seconds, they live forever. All can be thought of as made of three quarks in just 3 flavours, with the  {\it up} and {\it down} quarks essentially degenerate in mass, and the strange quark 120-150 MeV heavier. The same quarks, combined with the corresponding antiquarks, make the ground state, pseudoscalar, mesons: $\pi, K, \eta$ and $\eta'$. In the quark model, these states are simply  ${\overline q}q$  with spin, $S$, equal to zero, and with no orbital angular momentum, $L$. Their $S=1$ companions form the vector multiplet: the $\rho$, $\omega$, $\varphi$  and $K^*$'s. Their strong decays into pseudoscalar mesons provide the clues to their identities. They decay by creating a ${\overline u}u$ or ${\overline d}d$ pair from the vacuum. Then the $K^*$'s naturally decay to $K\pi$, the $\rho$ to $\pi\pi$. The degeneracy in mass of the $\omega$ to the  isotriplet $\rho$ suggests it too is built of {\it up} and {\it down} quarks. The mass differences of the $\varphi$ and $K^*$, and the $K^*$ and $\rho$, hint the $\varphi$ is largely ${\overline s}s$. The proof is provided by the fact the $\varphi$ decays to ${\overline K}K$, when it has a far larger phase-space to decay, like the $\omega$, to $3\pi$. Decays provide the \lq\lq flags" that tell us the make-up of these states. 
\begin{figure}[t]
\centerline{%
\includegraphics[width=12.5cm]{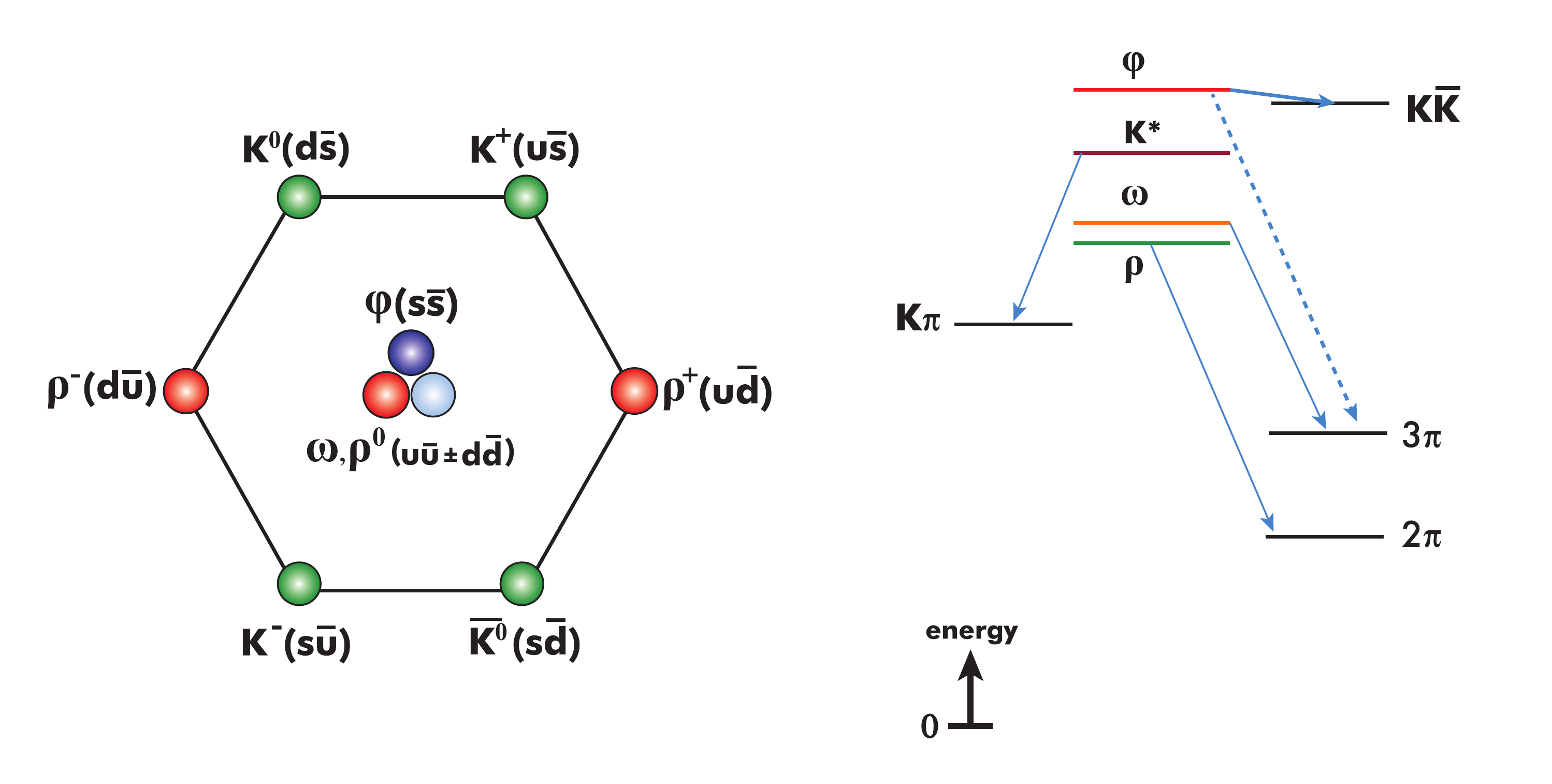}}
\caption{The ideal quark model multiplet on the left is a good approximation for the 9 lightest vector mesons, consistent with the decay pattern shown on the right.}
\end{figure}

The fact that these, like all excited mesons, decay means they are not simply ${\overline q}q$ systems. Their Fock space must contain at least four quark components, that rearrange themselves into two lighter mesons. For the lightest vectors, these components are small and do no more than move the pole in their propagators from the real energy axis of stable particles into the complex plane, reflecting decay. This movement is small because of the $P$-wave nature of their coupling to two pseudoscalars. Consequently, ${\overline q}q$ components dominate, as depicted in the upper graphs of Fig.~2. This appears to be the case for the highest spin states at any given energy, that is for those states lying along the leading Regge trajectories. However, the proportion of ${\overline q}q$ and hadronic modes is different in mesons with other quantum numbers. 

Eef van Beveren recognised that if the degrees of freedom of each meson are not just ${\overline q}q$ but its hadronic decay channels too, dynamics could naturally generate orthogonal states in which the hadron modes would bind. Indeed, such states would not then be pure molecules, but have some residue of their ${\overline q}q$ seeds, as illustrated in the lower graphs of Fig.~2. The binding of such states is a matter of dynamics~\cite{eef1,lutz}. If the coupling is $S$-wave then the hadronic components are most likely to bind. Eef and collaborators~\cite{eef2} thought the lightest scalar mesons might well be of this type. Others, like myself, took some time to realise the importance of the hadronic degrees of freedom~\cite{tornqvist,mrp-boglione}, as we will discuss below. What has highlighted this to the world in general are experimental developments in the charmonium sector.
\begin{figure}[t]
\centerline{%
\includegraphics[width=11.7cm]{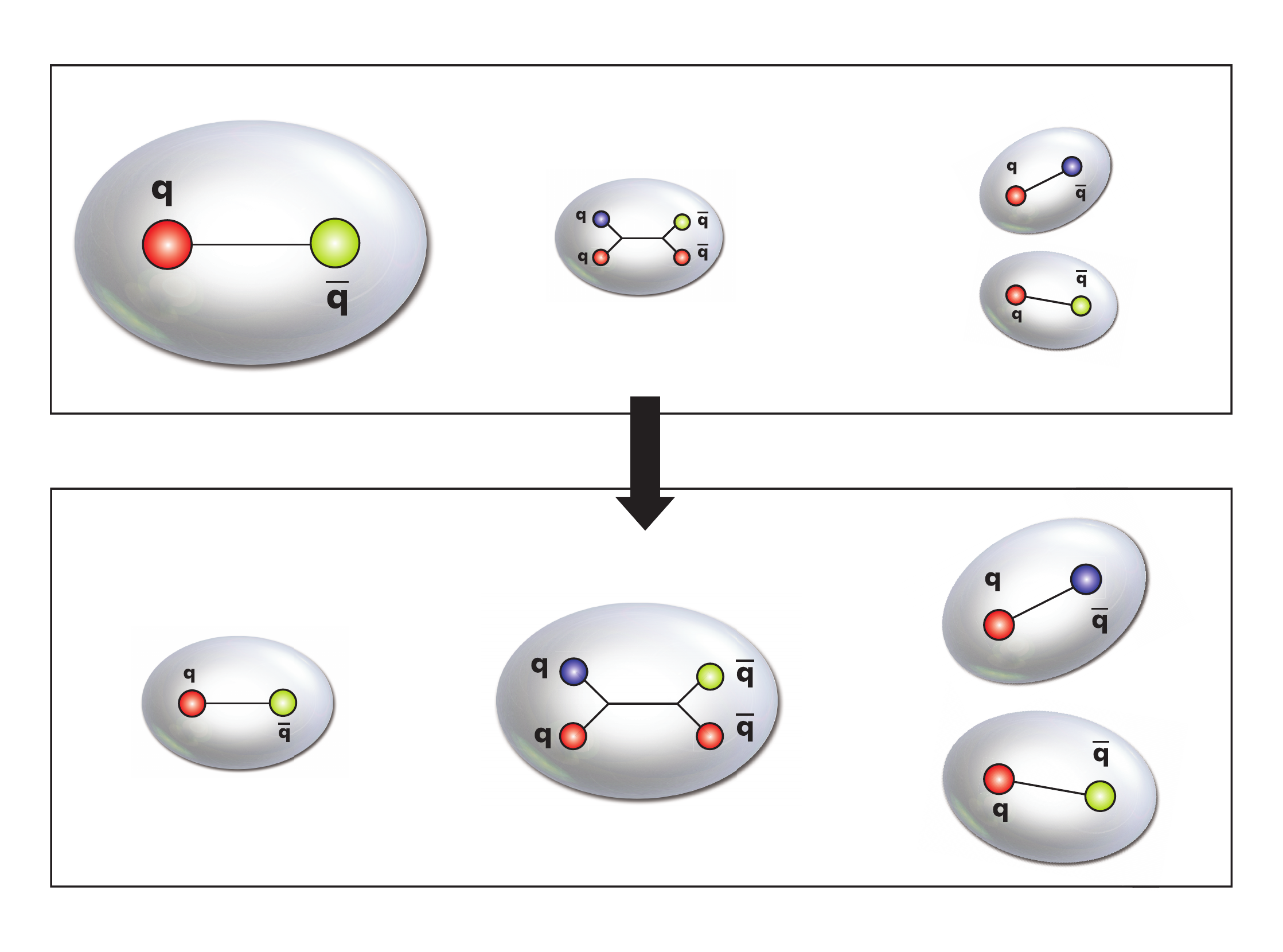}}
\caption{The Fock space of meson states including not just ${\overline q}q$ components, but also four quark and two or more meson degrees of freedom. The upper figure represents those dominated by  ${\overline q}q$ configurations, like the well-known vector and tensor mesons. The lower figure represents possible orthogonal states that with $S$-wave coupling to hadronic components might in the right dynamical situation be dominated by the binding of these degrees of freedom.}
\end{figure}

The discovery forty years ago of the $J/\psi$, followed shortly by the $\psi'$, quickly led to simple potential models of the emerging charmonium spectrum. Such models flourished even more with the later  discovery of the still heavier bottomonium sector. With the addition of relativistic corrections for the lightish charm quark, the whole spectrum of ${\overline c}c$ states has long been predicted, Fig.~3. The vector states can be found in $e^+e^-$ collisions. As the energy increases above ${\overline D}D$ threshold,  the narrow $1^{--}$ states give way to wider $\psi$'s that are heavy enough to decay to states with naked charm. However, even the $\psi'''$ is found to be not quite where it is predicted, as are many other states too. This hinted that the opening of decay channels shifted their masses~\cite{eichten}. This was no surprise to Eef.  Several calculational schemes for these shifts have been developed~\cite{eichten,barnes-swanson, wilson}, with reasonable success. What did surprise the wider community was that there could be orthogonal states in which hadron channels were the dominant component of a meson.      

\begin{figure}[b]
\centerline{%
\includegraphics[width=9.cm]{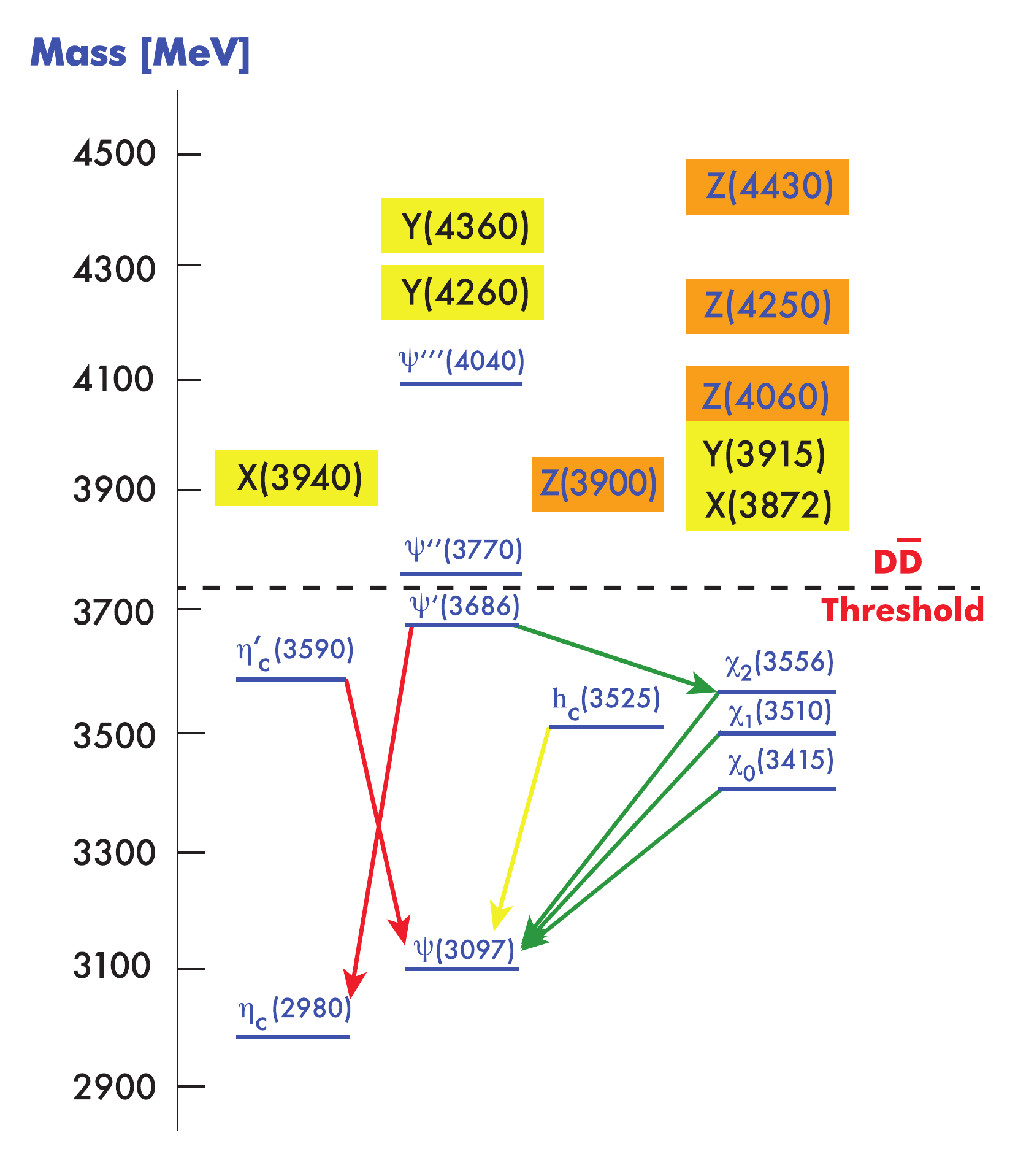}}
\caption{The observed charmonium spectrum with the more recently discovered {\it unexpected} $X,\ Y,\ Z$ states.}
\end{figure}

\begin{figure}[t]
\centerline{%
\vspace{-4mm}
\includegraphics[width=8.0cm]{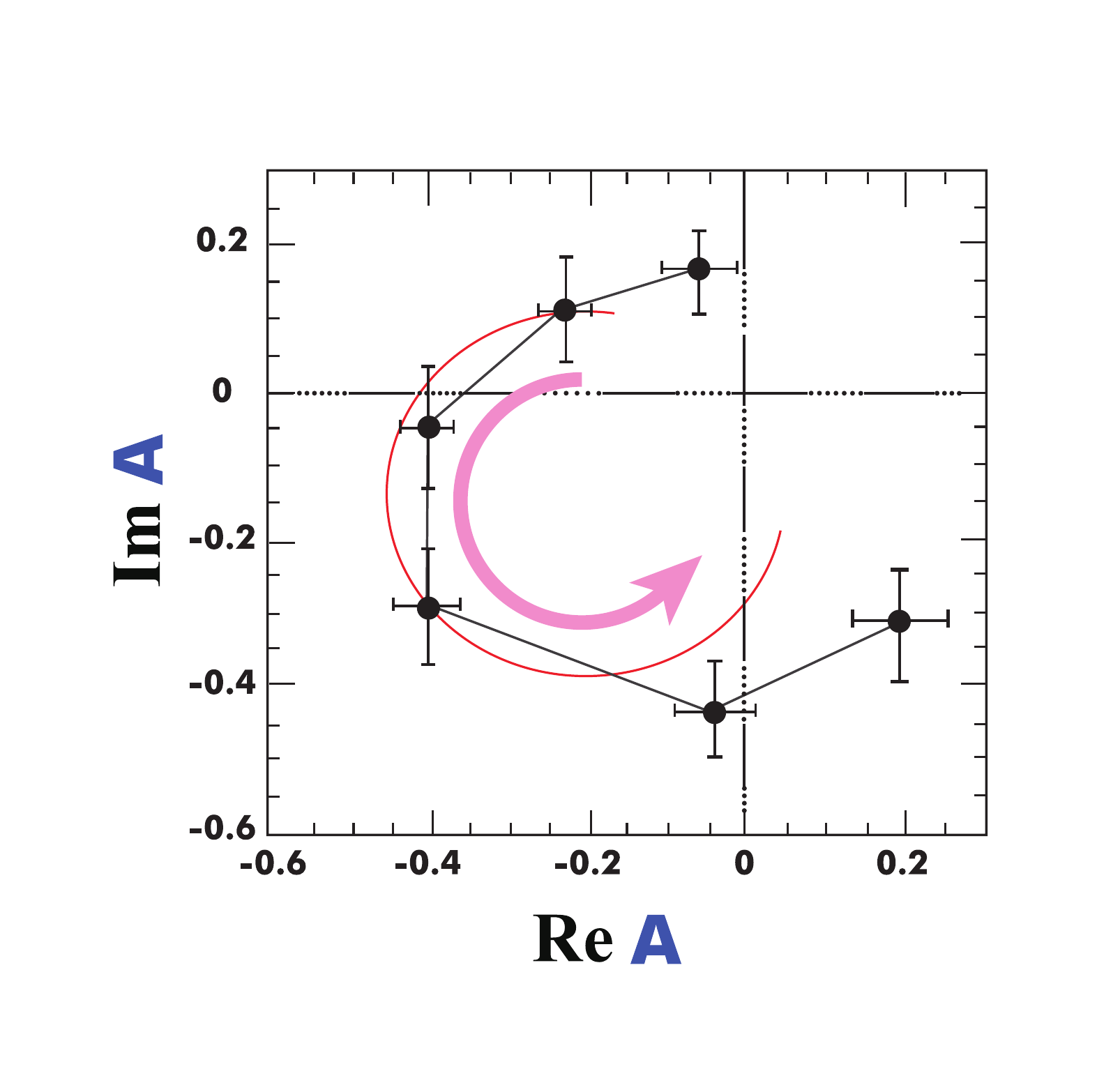}}
\vspace{-4mm}
\caption{Argand plot of the $Z_c$ amplitude (in arbitrary units) in six bins of $M(\pi\psi')^2$ as found by LHCb~\cite{ZcLHCb}. The smooth curve is the result of a Breit-Wigner fit with mass 4475~MeV and width 172~MeV. The phase is relative to the helicity-zero $K^*(890)$, which  is taken to be real.}
\end{figure}

\section{Excited spectrum: the $X, Y, Z$ mesons}
 
The $X(3872)$ was the first of what we now know is a whole series of unexpected $X, Y, Z$  states with hidden charm~\cite{XYZ}. The $X(3872)$ is found in $B\to K X$, where the $X$ is observed in the $J/\psi\pi\pi$ spectrum. Even though its mass is 130-140 MeV above ${\overline D}D$ thresholds, it has a width of at most 2.3 MeV. What makes a state with a mass of  nearly 4~GeV live 50 times longer than expected? The $X(3872)$ has been found in many experiments and now confirmed to have  $J^{PC}=1^{++}$ quantum numbers by LHCb. It has a very close $S$-wave connection to ${\overline D^{*0}}D^0$ threshold, and is so narrow that it has no overlap with the corresponding charged ${\overline D^*}D$ channel, which is 8~MeV heavier. This could well be one of Eef van Beveren's anticipated states, orthogonal to a host of charmonium states all with charmed-anticharmed hadronic modes: ${\overline D}D$, ${\overline D^*}D$, ${\overline D^*}D^*$, ${\overline D_s}D_s, \cdots$, and predominantly ${\overline D^{*0}}D^0$ and ${\overline D^0}D^{*0}$~\cite{TornqX}. Incidentally, analysis by Susana Coito, van Beveren and Rupp~\cite{coito} showed this state can't be  purely a hadronic molecule. Of course, if its seed is a ${\overline c}c$ state, then a residual component would be expected to remain (Fig.~2).

As illustrated in Fig.~3, a whole host of {\it unexpected} states have been discovered in recent years, the $X, Y, Z$ states. While all around 4~GeV are connected to charmonium, the most spectacular is the $Z_c^\pm(4340)$, seen in $B\to K^\mp Z_c^\pm$ where $Z_c^\pm$ appears as a \lq\lq peak'' in the $\pi^\pm \psi'$ spectrum as found by Belle. This strong decay of the charged state  tells us it must contain more than a ${\overline c}c$, with a $u{\overline d}$ or $d{\overline u}$ depending on its charge. Whether it is a four quark state, or a hadronic molecule of a charmed and anti-charmed meson awaits further examination. This state has been studied by LHCb~\cite{ZcLHCb}, who found its amplitude has the phase variation expected of a resonance, Fig.~4. Nevertheless, a less model-dependent Dalitz analysis of the $B$-decay to $K\pi\psi'$ is required to be certain. Moreover, further analyses are essential to confirm each of the $X, Y, Z$ states: bumps don't equal hadrons, only poles in the complex energy plane do. Tantalizingly all the present signals hint at a strong $S$-wave coupling to nearby hadronic decay channels. Moreover, these states seem in turn to be connected through further decays to each other, {\it e.g.} $Y(4260)\to \gamma X(3872)$~\cite{YtoX}. There are also indications that such $X, Y, Z$ states  do not just come in the charmonium, but in bottomonium and strangeonium sectors too, for instance the series of $Z_b\to \Upsilon(nS)\pi$ (with $n= 1,2,3,\cdots$)~\cite{Zb} and $Y(2175) \to \varphi f_0(980)$~\cite{Y2175}. 

The discovery of the $X, Y, Z$ mesons, together with novel charmed states among the $D$ and $D_s$ mesons, has revitalised interest in spectroscopy, and a whole series of new experiments in
hadroproduction, photoproduction, $e^+e^-$ and ${\overline p}p$ annihilations are planned to study these further and perhaps discover yet more states.   

\section{Light scalars: bound by interquark or interhadron forces}

It is in the light hadron sector, as Eef has long known, that intimations of multi-quark or hadronic molecules are there~\cite{eef2,jaffe,tetra,wi}. For some decades it has been understood that the 9 lightest scalars in Fig.~5 from the PDG tables~\cite{pdg} do not fit the $L=S=1$ ${\overline q}q$ states expected in the simple quark model. Rather the heavier $a_0(1450)$, $K_0^*(1430)$, together with some mixture of the $f_0(1370)$, $f_0(1500)$ and $f_0(1710)$ better fit this bill.
But then what are the lower lying scalars: four quark or hadronic molecules? The near degenerate mass of the isotriplet $a_0(980)$ and isosinglet $f_0(980)$,  both coupling strongly to the nearby ${\overline K}K$ channels is fulfilled whether they are tetraquark ${\overline {sn}}sn$ states (with $n = u, d$), or largely ${\overline K}K$ molecules~\cite{achasov_kk,coito_scalars}. Can one tell the difference?
\begin{figure}[t]
\centerline{%
\includegraphics[width=7.3cm]{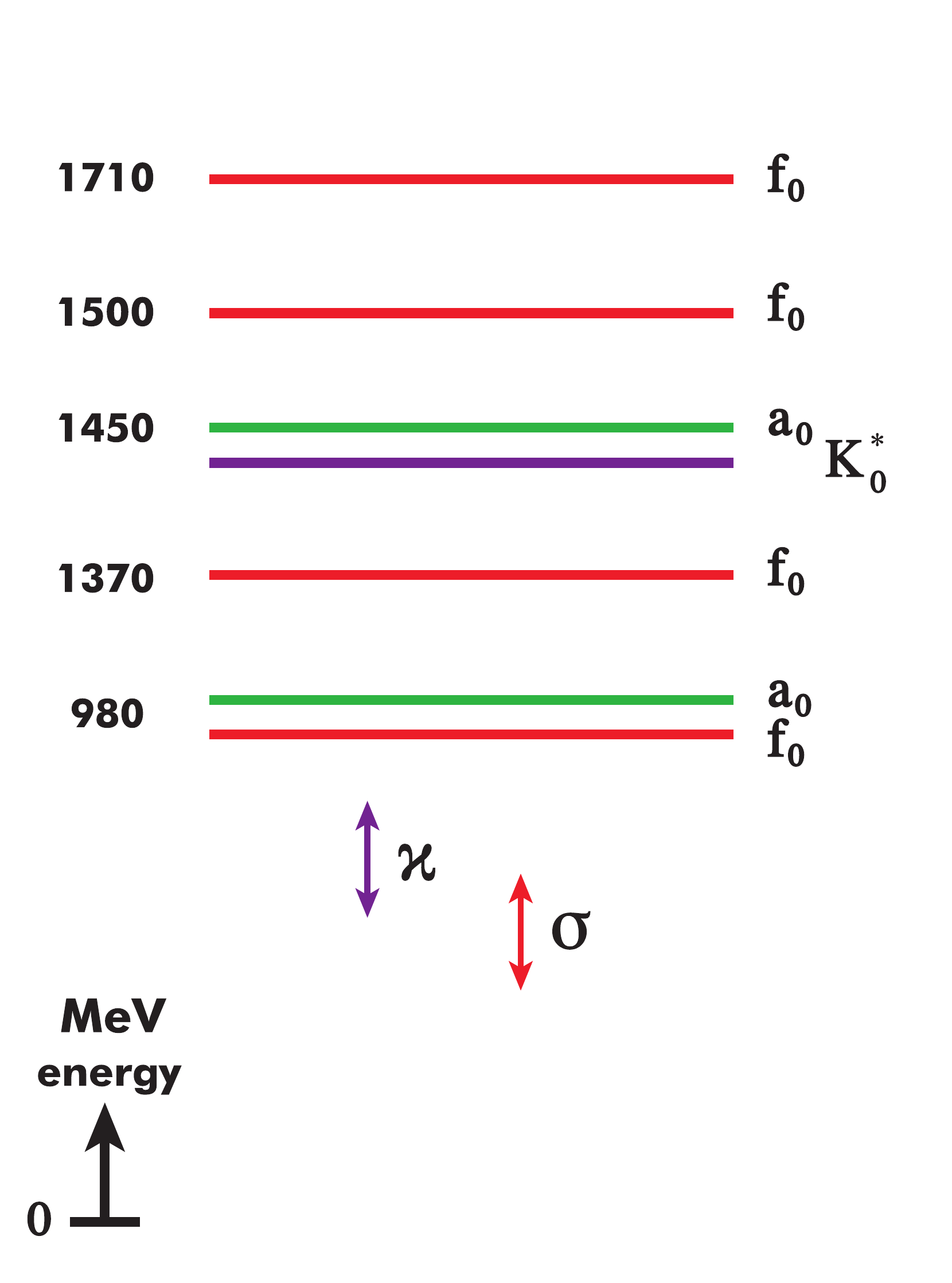}}
\caption{The observed spectrum of scalar mesons~\cite{pdg}. The arrows represent the \lq\lq mass'' of the particularly broad isosinglet $\sigma$ and isodoublet $\kappa$ mesons.}
\end{figure}

The issue of whether a state is bound by interquark forces or interhadron forces is a matter of the range of the interaction. Indeed, it is this that allows us to know that the deuteron is  a bound system of a proton and a neutron, and not a six quark bag. The way to study this was presented by Weinberg~\cite{weinberg}. These arguments were recast for the meson sector, when a state is close to an inelastic threshold, by my long time collaborator David Morgan~\cite{morgan}, who demonstrated a pole counting \lq\lq theorem''. A molecule is dominated by the pole in the complex energy plane corresponding to a bound state, while a state dominated by interquark forces, whether ${\overline q}q$ or tetraquark, also acquires poles on other sheets (see also \cite{baru}). Of course, the world is not ${\overline q}q$ or molecular, but a mixture of both degrees of freedom. The deuteron being predominantly a bound state of a proton and a neutron, does not mean it has no 6-quark configurations. As illustrated by the lower graphs in Fig.~2, molecular states may in reality be seeded by ${\overline q}q$ components. While these configurations may be small they are not zero. Consequently, the pole counting question is not \lq\lq is there one pole or two anywhere in the complex plane?'', but rather \lq\lq is there is one or two {\it nearby}?'': {\it near} in momentum being inversely related to the range of the forces that do the binding~\cite{wi,morgan,hanhart}.

\begin{figure}[b]
\vspace{-7mm}
\centerline{%
\includegraphics[width=9.3cm]{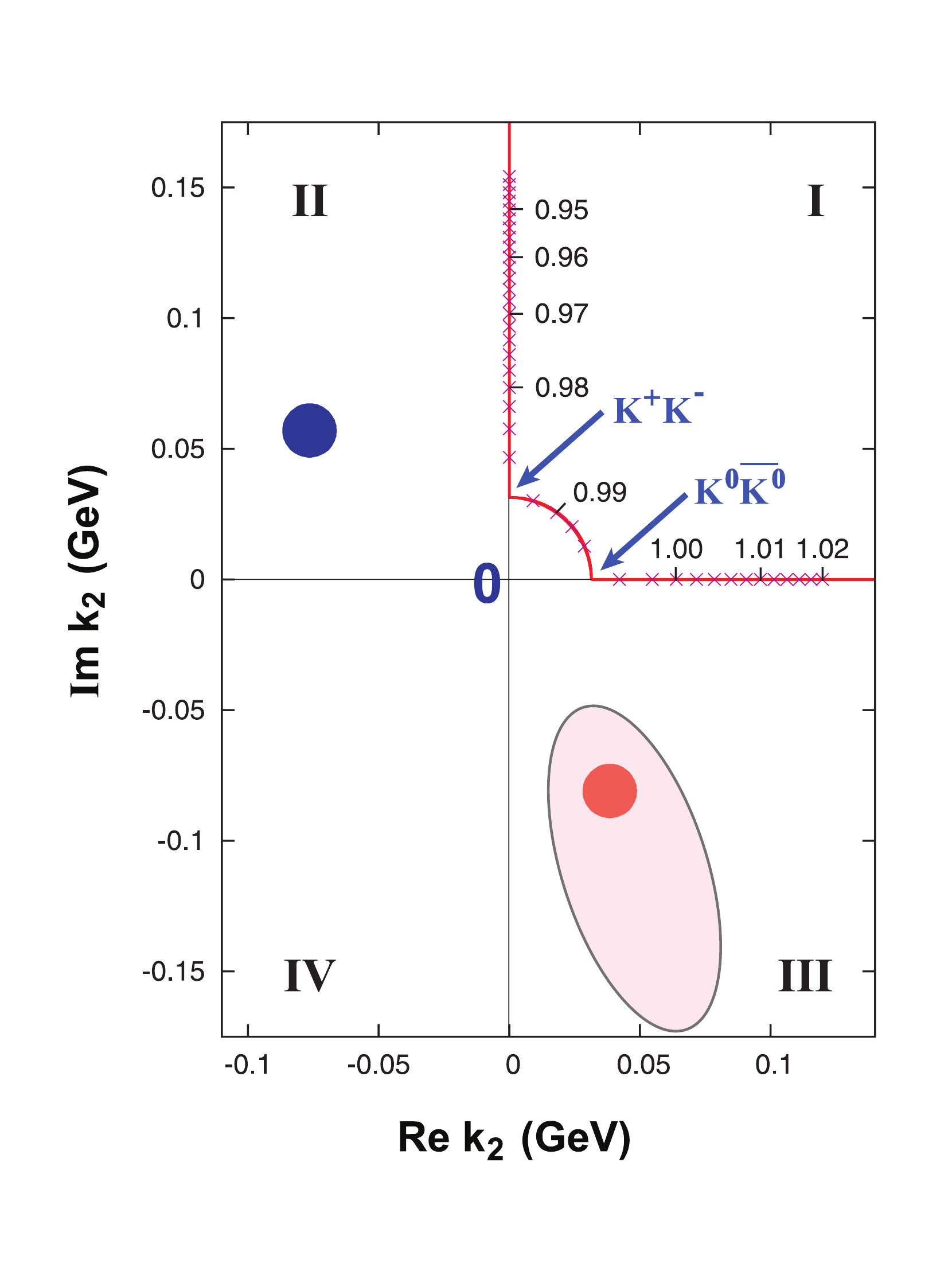}}
\vspace{-6mm}
\caption{The complex $k_2$-plane in the neighbourhood of the two ${\overline K}K$ thresholds, where $k_2=\frac{1}{2}\sqrt{s-4m_{Kc}^2}\,+\,\frac{1}{2}\sqrt{s-4m_{Kn}^2}$ is the mean ${\overline K}K$ c.m. 3-momentum with $K_c$ and $K_n$ the charged and neutral kaon masses, respectively. The Roman numerals label the four sheets.
 The c.m. energy, $\sqrt{s}$, is marked ($\times$) every 2~MeV, with the energy in GeV enumerated every 10~MeV. The circles indicate the position of poles on Sheets II and III. The shaded region is explained in the text~\cite{wilson_f0}.} \label{fig:k2}
\end{figure}

In the case of the $f_0(980)$ that couples to $\pi\pi$ and ${\overline K}K$, the complex $s$-plane has 4-sheets corresponding to choosing the signs of the square roots of the $\pi\pi$ and ${\overline K}K$ centre-of-mass 3-momenta, $k_i$, where $i=1$ labels the pion channel and $i=2$ the kaon one. This follows from the fact that unitarity requires new contributions to the discontinuity of a partial wave amplitude proportional to each $\rho_i = 2k_i/\sqrt{s}$. These sheets are usefully separated by considering the $k_2$-plane, shown in Fig.~6.  If all kaons had the same mass, then ${\overline K}K$ threshold would be at the origin. The 8~MeV difference of charged and neutral kaon pairs separates the thresholds, Fig.~6. This mass difference is assumed to be the sole source of isospin breaking. 

A Breit-Wigner representation of the $f_0$-amplitude automatically has two poles, being a function of $s$ (and so quadratic in $k_2$). Consequently to answer the question how many poles are nearby requires a more flexible representation respecting unitarity and analyticity in the $k_2$-plane. This is provided by Jost functions~\cite{dm-mrp}. Using these, the amplitudes for $\pi\pi\to\pi\pi$ and $\to {\overline K}K$ are represented by functions in which the number of poles is specified. Data along the axes shown in Fig.~6, where the values of $\sqrt{s}$ are labelled every 10~MeV, with crosses between indicating  2~MeV steps, is where experiment is performed. When the classic meson-meson scattering data in 20~MeV bins are fitted for the $I=J=0$ partial wave, there is always a pole on sheet II, with a  location that is well-defined, as in Fig.~6. Whether there is a pole or not nearby on sheet~III, the quality of fit cannot distinguish~\cite{dm-mrp}. We would need data on meson scattering of much greater precision to achieve that, but no new hadron peripheral production experiments are planned.

Fortunately, the same state can be accessed in heavy flavour decays. While the $f_0(980)$ produces a dramatic dip in the $I=J=0$ component of the elastic $\pi\pi$ cross-section, it creates a peak in the $\pi\pi$ spectrum seen in the decays $J/\psi\to\varphi\pi^+\pi^-$ and $D_s^+\to \pi^+\pi^+\pi^-$, reflecting its strong coupling to hidden strangeness. The data on $D_s\to \pi(M^+M^-)$ decay from BaBar have been partial wave analysed to extract both the $I=J=0$ $M =\pi$ and $K$ amplitudes,  their moduli and phases~\cite{babar1,babar2}. What is more the kaon pair data are in 4 MeV bins and constrain the simultaneous fit most precisely. Nevertheless, these $D_s$ amplitudes still allow both a one pole and two pole fit of equal quality. However, the quality of the two pole fit deteriorates rapidly~\cite{wilson_f0}, if the pole on sheet~III is outside the shaded region shown in Fig.~6. Careful scrutiny shows that when the pole is in that region, its residues (both coupling to $\pi\pi$ and ${\overline K}K$) are much much smaller than those of the pole on sheet~II. Thus a two pole fit is only possible, when the second pole is essentially not there. Thus data along the real energy axis are sufficient to conclude the $f_0(980)$ behaves as if it were a hadronic molecule. The pole on Sheet~II is all that matters. If it were a true bound state of the ${\overline K}K$ channel then the pole would be on the imaginary axis in Fig.~6. That the state decays to $\pi\pi$ moves the pole away to where it is shown in Fig.~6, but this does not acquire the nearby companion sheet~III pole that a state dominated by interquark forces requires~\cite{wilson_f0}.

Eef van Beveren, George Rupp and coworkers anticipated this long ago~\cite{eef2}. With the scalar ${\overline q}q$ seeds  up near the tensor multiplet at 1.2-1.5~GeV, the strong $S$-wave coupling of these scalars to channels with pseudoscalar pairs would not only move an ${\overline s}s$ seed, for instance, into the complex energy plane as required for a fully fledged hadron, but  a second state strongly coupled with, and close to, ${\overline K}K$ threshold would result, Fig.~7. The perfect illustration of what hadronic degrees of freedom do in Fig.~2. Dynamics would not just turn an ${\overline s}s$ seed into a decaying hadron, but generate a state that is largely a ${\overline K}K$ molecule too: a remarkable insight into what appears to be the truth. That the short-lived scalars the $\sigma$ and $\kappa$ are dominated by their $\pi\pi$ and $K\pi$ components becomes natural~\cite{mrp_scalars}. The fact that these states are above their corresponding thresholds means their binding is weaker and they decay  even more rapidly. Consequently, the two photon coupling of the $\sigma$ is dominated by its $\pi\pi$ configuration~\cite{mrp-prl,dai_gg}, regardless of whether it has a smaller ${\overline n}n$ or ${\overline nn}nn$ (or even gluonic) core. Though Belle data~\cite{belle_gg} now allow an accurate determination of the two photon coupling of the $f_0(980)$~\cite{dai_gg}, the predictions of what this should be for a molecule or simple ${\overline q}q$ states is as yet less reliable~\cite{hanhart_gg,achasov_gg,giacosa_gg}.

\begin{figure}[t]
\begin{center}
\includegraphics[width=0.74\textwidth]{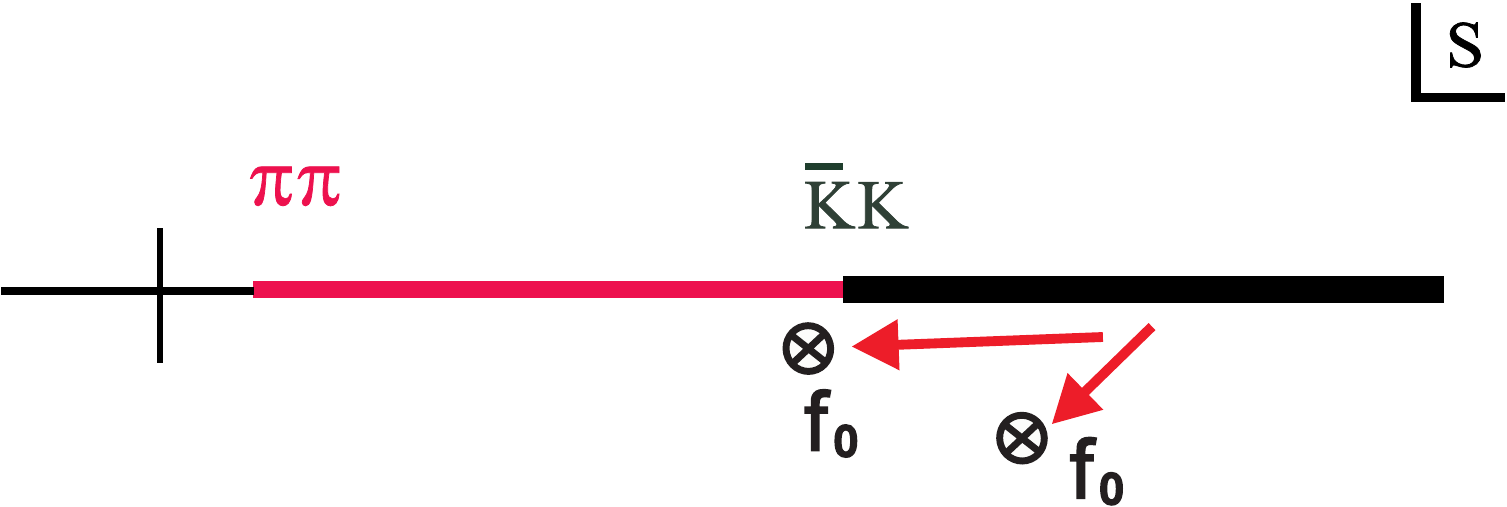}
\caption{The analytic structure of the $f_0$-propagator in the complex $s$-plane, where $s$ is the momentum squared. This has cuts at $\pi\pi$, $4\pi$, ${\overline K}K$, {\it etc.} thresholds. The ${\overline s}s$ \lq\lq seed'', for instance, is a pole on the real axis. When the ${\overline{K}}K$ channels are switched on, this pole moves onto the nearby unphysical sheet as arrowed. In the calculations of~\cite{eef2}, a second pole is dynamically generated close to ${\overline{K}}K$ threshold. }
\end{center}
\vspace{-0.5cm}
\end{figure}

Data of sufficient precision on $X(3872)\to D^0{\overline D^{*0}}$, $D^+{\overline D^{*-}}$, and $J/\psi\pi^+\pi^-$ might allow a similar analysis~\cite{dai_zheng} to that described here for the $f_0(980)$, and for the other $X,Y, Z$ states too. While the poles of the $S$-matrix define the spectrum of hadrons, and their position in the complex energy plane are process independent, it is their couplings in production and decay that teaches us about the way the dynamics of QCD works. Lattice calculations are also starting  to include the effect of hadronic channels, and so learn about the way they influence the masses and properties of states. Of course, decay channels become more important as the mass of the pion advances towards its physical value. 

The long range aspects of QCD encode confinement. Understanding this is critical even at LHC energies. For though there one studies hard interactions at scales a thousand times smaller than the size of a hadron, and with times of only $10^{-26}$ of a second, to get down to those scales one has to understand how the protons that collide break up into the tiny entities that interact (current quarks, gluons and their possible {\it super}-partners) and importantly how after collisions these get back to make hadrons, protons, pions, kaons, {\it etc.}, observed in detectors. At the LHC these long distance interactions are modelled  in the Monte Carlo generators. By recognising that the whole life cycle of hadrons is essential to their existence, Eef van Beveren has pointed the way to gaining a better understanding of colour confinement and its consequences. The thirst for knowledge of how QCD really works remains unquenched.
 It is only with a detailed knowledge of  the strong interaction web that surrounds the femto-universe that we can peer in and untangle what is truly beyond the Standard Model. 

\vspace{3mm}
  It is pleasure to thank Eef van Beveren, whose research achievements and seventieth birthday we celebrate at this fascinating workshop. I wish Eef a long, happy and healthy retirement. I am grateful to George Rupp for organizing this meeting, and arranging support for  my visit to Coimbra.
The work was authored in part by Jefferson Science Associates, LLC under U.S. DOE Contract No. DE-AC05-06OR23177. 
\vspace{3mm}

\end{document}